\documentclass[twocolumn,preprintnumbers,amsmath,amssymb,reprint,groupedaddress,amsmath,citeautoscript,flushbottom,floatfix,superscriptaddress]{revtex4}
\usepackage{graphicx}
\usepackage{dcolumn}
\usepackage{bm}
\usepackage[colorlinks=true, linkcolor=blue, citecolor=blue, urlcolor=blue]{hyperref}

\newcommand{\vc}[1]{\boldsymbol{#1}}

\begin{document}

\title{Ultrasensitive strain modulation of terahertz magnons at a magnetic phase transition}

\author{Lichen~Wang}
\email{lichen.wang@fkf.mpg.de}
\affiliation{Max Planck Institute for Solid State Research, Heisenbergstrasse 1, D-70569 Stuttgart, Germany}
\author{Sajna~Hameed}
\affiliation{Max Planck Institute for Solid State Research, Heisenbergstrasse 1, D-70569 Stuttgart, Germany}
\author{Yiran~Liu}
\affiliation{Max Planck Institute for Solid State Research, Heisenbergstrasse 1, D-70569 Stuttgart, Germany}
\author{Manuel~Knauft}
\affiliation{Max Planck Institute for Solid State Research, Heisenbergstrasse 1, D-70569 Stuttgart, Germany}
\author{Kazuki~Higuchi}
\affiliation{Max Planck Institute for Solid State Research, Heisenbergstrasse 1, D-70569 Stuttgart, Germany}
\author{Maximilian~Krautloher}
\affiliation{Max Planck Institute for Solid State Research, Heisenbergstrasse 1, D-70569 Stuttgart, Germany}
\author{Sonia~Francoual}
\affiliation{Deutsches Elektronen-Synchrotron DESY, Notkestrasse 85, D-22607 Hamburg, Germany}
\author{Giniyat~Khaliullin}
\affiliation{Max Planck Institute for Solid State Research, Heisenbergstrasse 1, D-70569 Stuttgart, Germany}
\author{Huimei~Liu}
\email{huimeiliu@nju.edu.cn}
\affiliation{Max Planck Institute for Solid State Research, Heisenbergstrasse 1, D-70569 Stuttgart, Germany}
\affiliation{National Laboratory of Solid State Microstructures and School of Physics, Nanjing University, Nanjing 210093, China}
\affiliation{Collaborative Innovation Center of Advanced Microstructures, Nanjing University, Nanjing 210093, China}
\author{Matteo~Minola}
\email{m.minola@fkf.mpg.de}
\affiliation{Max Planck Institute for Solid State Research, Heisenbergstrasse 1, D-70569 Stuttgart, Germany}
\author{Bernhard~Keimer}
\email{b.keimer@fkf.mpg.de}
\affiliation{Max Planck Institute for Solid State Research, Heisenbergstrasse 1, D-70569 Stuttgart, Germany}

\date{\today}
\flushbottom

\begin{abstract}
Antiferromagnets typically host spin-wave (magnon) excitations in the terahertz (THz) regime, offering a promising platform for high-speed magnonic information technologies. Harnessing these excitations requires sensitive control of their spectral properties. Here we use resonant x-ray diffraction and Raman scattering to demonstrate uniaxial-strain control of the antiferromagnetic (AFM) ground state and THz magnon excitations in the layered Mott insulator Ca$_2$RuO$_4$. Although the states separated by the strain-induced phase transition differ only by the sign of the weak and partially frustrated interlayer interaction, their magnon energies differ by more than 10\% ($\sim$ 0.3 THz). Our theoretical analysis explains this surprising observation by tracing the origin of both the sign reversal of the interlayer coupling and the magnon energy to the spin-orbital composition of the Ru valence electrons. The extreme strain sensitivity of the THz magnon energy near a magnetic phase transition opens up pathways towards a new generation of ``transition-edge'' magnonic devices.
\end{abstract}
\maketitle

Driven by the prospect of using magnons to transmit information with much lower dissipative losses than charge carriers in conventional electronics, the past two decades have seen tremendous progress in the controlled manipulation of gigahertz (GHz) ferromagnetic magnons and in the realization of ``magnonic'' device concepts \cite{Fle24}. Recent research has begun to investigate the generation, propagation, and detection of magnons in antiferromagnets, which do not generate parasitic stray fields and typically exhibit frequencies in the THz range, thus opening up perspectives for nanoscale, high-speed magnonic devices \cite{Fle24}. In analogy to field-effect devices in conventional electronics, current exploratory efforts aim to generate magnon potentials that are tunable on-demand. Possible approaches include electric-field modulation of ``electro-magnons'' in multiferroic materials \cite{Fle24}, and modulation of THz magnons in antiferromagnets by uniaxial strain from piezo-electric actuators \cite{Kim22}. 

Here we show that the sensitivity of these modulations can be greatly enhanced in the vicinity of strain-induced magnetic phase transitions. Such transitions have recently been reported in various quasi-two-dimensional (quasi-2D) magnets under different strain variants, including uniaxial strain modulation of the semiconducting van-der-Waals compound CrSBr with antiferromagnetically coupled ferromagnetic layers \cite{Cen22} and of the insulating triangular-lattice antiferromagnet PdCrO$_2$ \cite{Sun21}, as well as biaxial \cite{Shr22} or shear \cite{Sha22} strain manipulation of the square-lattice antiferromagnet Sr$_2$IrO$_4$. In all of these systems, the phase transition separates different stacking patterns of magnetic layers with exchange interactions whose strength greatly exceeds those of the interactions between the layers. As the mechanism underlying these transitions is thought to involve strain modulation of the weak interlayer exchange interactions, the difference between the magnon energies in both states is expected to be small. In line with this expectation, recent experiments on CrSBr field-effect devices have detected shifts of the magnon energy in the GHz regime \cite{Ste25}.

\begin{figure*}
        \centering{\includegraphics[width=1\textwidth]{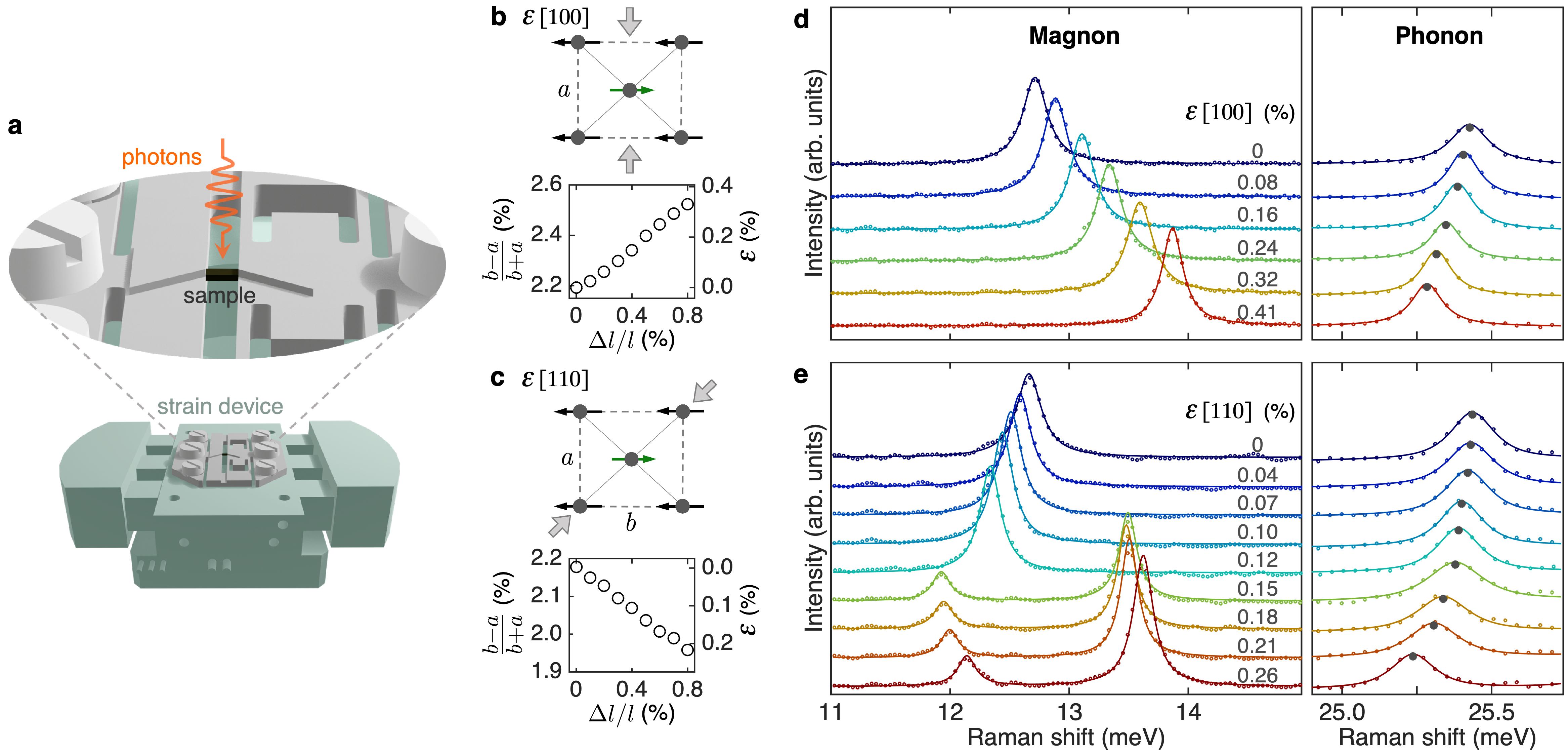}}
	\caption{\bf{}Uniaxial strain setup, structural response and Raman spectra under compressive strain along the [1 0 0] and [1 1 0] directions. a\rm{}, Schematic of the strain setup, see details in Methods and Supplementary Note 1. \bf{}b\rm{}, Top: schematic of strain application along the [1 0 0] direction. Lattice constants $a$ and $b$ correspond to the next-nearest Ru-Ru directions ($b > a$), with ordered pseudospins aligned along the [0 1 0] axis. Black circles represent Ru ions; black and green arrows indicate the antiparallel alignment. Bottom: lattice orthorhombicity $\frac{b-a}{b+a}$ and strain $\varepsilon$ under normalized compression $\Delta l/l$ along the [1 0 0] direction, where $\Delta l$ is the voltage-controlled piezoelectric displacement and $l$ is the sample length. Lattice orthorhombicity increases under compression along the [1 0 0] direction. \bf{}c\rm{}, Top: schematic of strain application along the [1 1 0] direction. Bottom: reduced orthorhombicity and corresponding strain $\varepsilon$. \bf{}d\rm{}, Raman spectra at 25 K under compressive strain along the [1 0 0] direction. Magnon spectra in the $B_\mathrm{1g}$ channel were acquired using cross linearly-polarized configuration for incident and scattered photons, while $A_\mathrm{g}$ phonon spectra were measured with parallel polarization geometry. Black dots mark the phonon energies. \bf{}e\rm, Raman spectra under compressive strain along the [1 1 0] direction. Additional Raman data and experimental details are provided in Supplementary Note 2 and Methods. } 
	\label{fig:1}
\end{figure*}

\begin{figure}
	\centering{\includegraphics[width=0.48\textwidth]{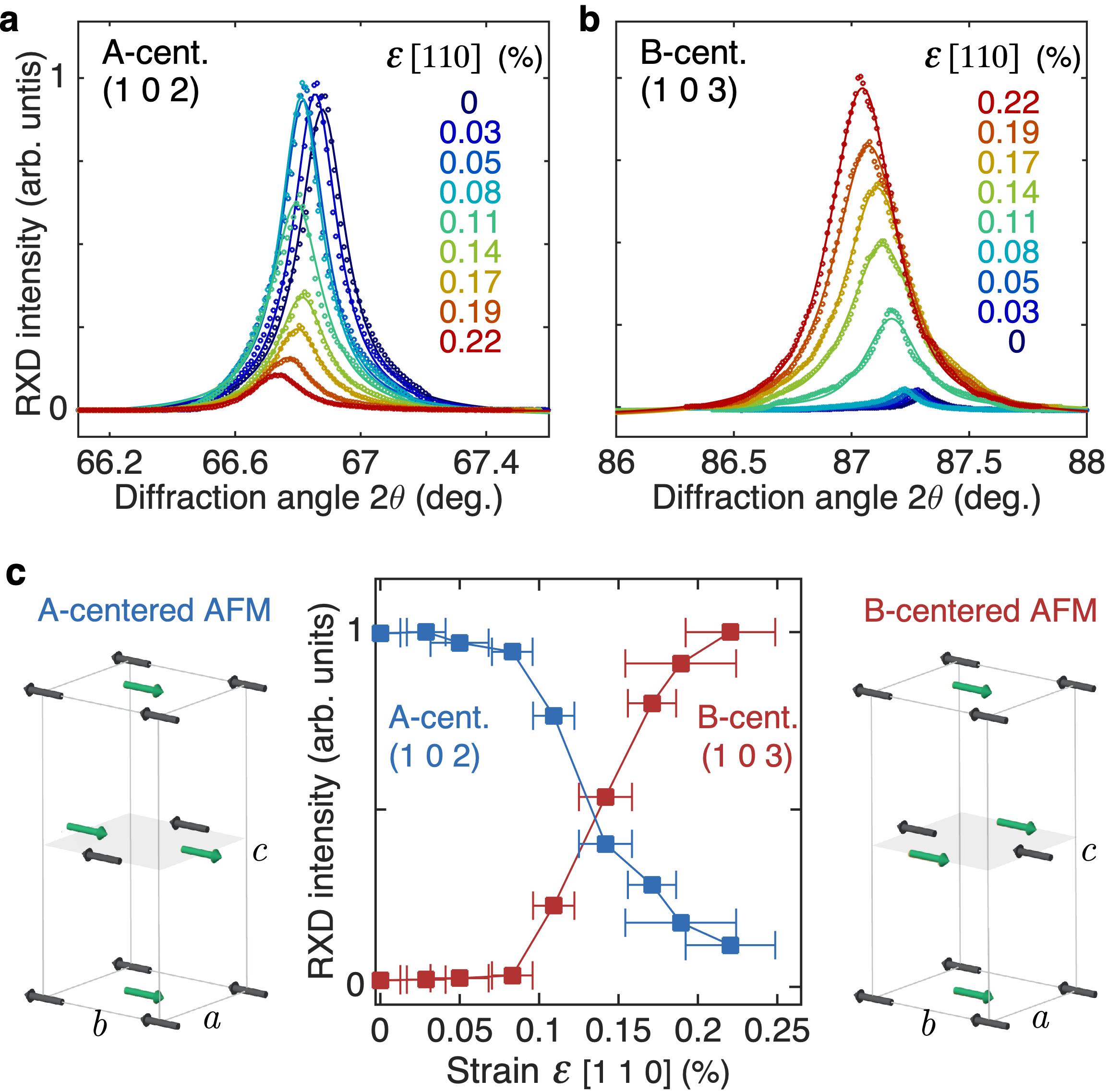}}
	\caption{\bf{}RXD results under compressive strain along the [1 1 0] direction and the magnetic phase transition. a--b\rm{}, Magnetic reflections of the A- and B-centered AFM orders measured at the Ru $L_3$ edge under varying strain at 20 K. The A-centered phase gives rise to magnetic reflections satisfying $h + l$ = odd and $k + l$ = even, while the B-centered phase corresponds to  $h + l$ = even and $k + l$ = odd. Longitudinal ($\theta-2\theta$) scans were performed with synchronized sample ($\theta$) and detector (2$\theta$) motion. \bf{}c\rm{}, Strain dependence of the integrated intensities for the A-centered (blue) and B-centered (red) magnetic reflections. The intensities for each phase are normalized to their maximum. Error bars indicate the strain uncertainty, derived from the simultaneous measurement of lattice constants (Supplementary Note 1). Schematics of the A- and B-centered magnetic structures are shown on the sides, with arrows in two colors representing antiparallel pseudospins of Ru ions.}
	\label{fig:2}
\end{figure}

We have performed resonant x-ray diffraction (RXD) and Raman scattering experiments on the quasi-2D square-lattice antiferromagnet Ca$_2$RuO$_4$ with N\'eel temperature $T_\mathrm{N} = 113$ K \cite{Bra98}. The Ru ions carry spin and orbital moments that can be described in terms of spin-orbit entangled pseudospins \cite{Tak21}. Since adjacent layers are stacked in a body-centered fashion (Fig.~\ref{fig:2}c), the interlayer exchange interactions between the pseudospins are nearly frustrated, and the net coupling between the layers is only stabilized by the small orthorhombic distortion of the tetragonal structure. In addition to the ``A-centered'' AFM structure observed under ambient conditions, prior experiments on pressurized, doped or off-stoichiometric Ca$_2$RuO$_4$ compounds with subtly different lattice structures have uncovered a ``B-centered'' AFM state with reversed sign of the interlayer coupling \cite{Bra98,Fri01,Ste05,Tan13,Pin18,Chi20,Por22}. Our experiments show that a transition from A-centered to B-centered structure can also be driven in pure Ca$_2$RuO$_4$ by $\sim$ 0.15\% uniaxial strain, analogous to the systems discussed above. Surprisingly, however, we found that this transition is associated with an abrupt, reversible jump of the zone-center magnon energy from $\sim$ 3.0 to 3.3 THz---orders of magnitude larger than the expected influence of the weak interlayer interactions ($J_c \sim 4\times10^{-3}$ THz \cite{Tre22,Kun15}). We present a quantitative theory that explains both the jump of the magnon energy and the sign reversal of the interlayer coupling in terms of strain-induced modulation of the spin-orbit admixture of the Ru pseudospins. Our discovery of a large difference of the magnon energies in nearly identical magnetic ground states separated by a sharp phase transition opens up perspectives for magnonic devices with interfaces whose transparency can be sensitively tuned by piezo-electric voltages.

The strain setup for Raman and x-ray scattering is shown in Fig.~\ref{fig:1}a. This system has been successfully used in several spectroscopic studies of correlated electron systems \cite{Kim18,Kim21,Kim22}. Compressive strain was applied along two in-plane directions: [1 0 0] (next-nearest Ru-Ru bonds) and [1 1 0] (nearest Ru-Ru bonds), as shown in Fig.~\ref{fig:1}b-c. The strain directly modifies the in-plane lattice orthorhombicity $(b-a)/(b+a)$, where $a$ and $b$ are the lattice constants along the orthorhombic axes ($b > a$). X-ray diffraction measurements show that the lattice orthorhombicity exhibits nearly linear but opposite responses to the two compression directions (Fig.~\ref{fig:1}b-c and Supplementary Note 1). Here, we model the evolution of magnon energy using an effective strain $\varepsilon = |(b-a)/(b+a)-\varepsilon_0|$, which captures the absolute change in lattice orthorhombicity from its intrinsic value $\varepsilon_0$. By this definition, $\varepsilon > 0$ represents compression for both strain directions. 

\begin{figure}
	\centering{\includegraphics[width=0.48\textwidth]{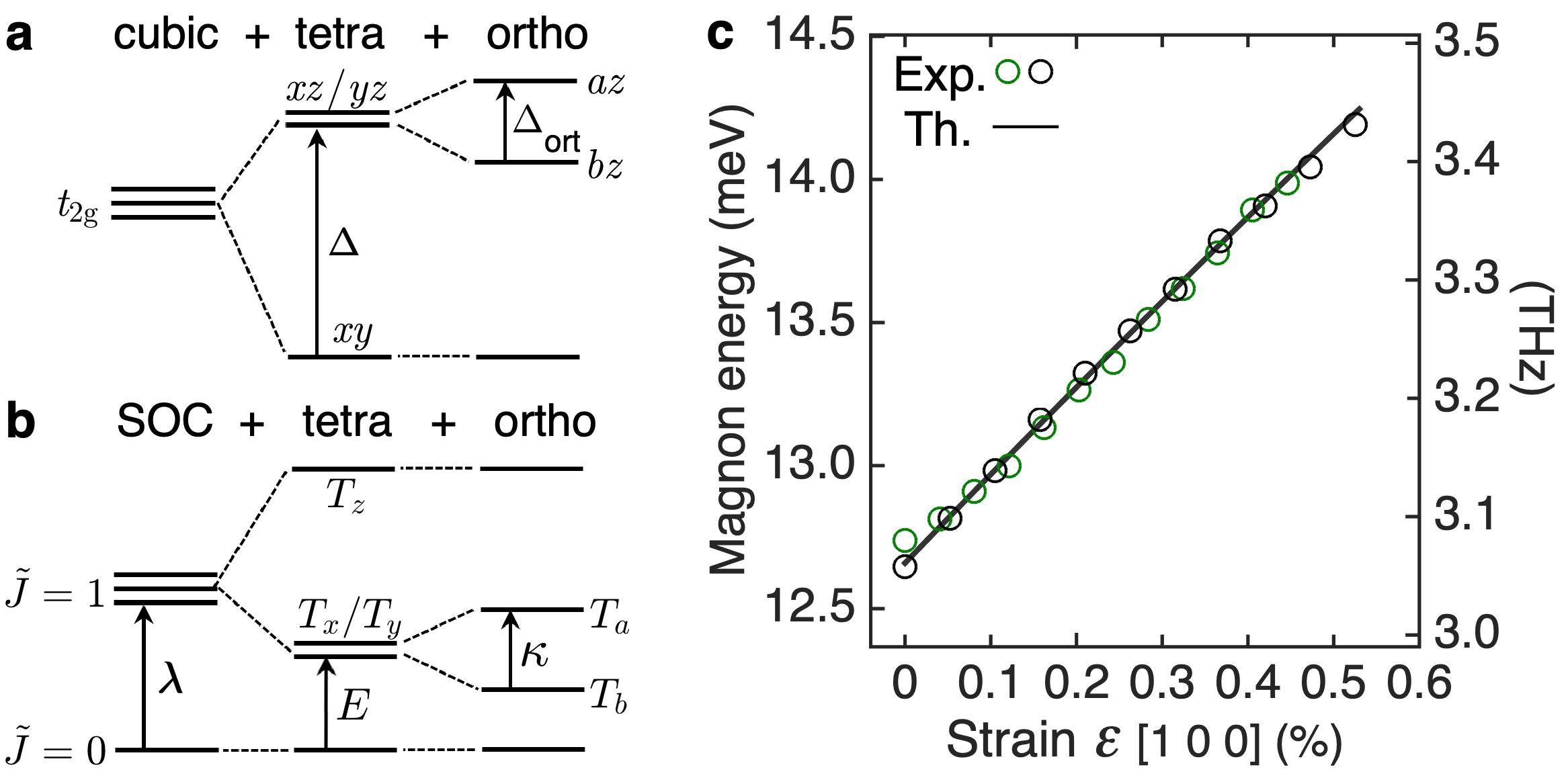}}
	\caption{\bf{}Electronic structure of $t_\mathrm{2g}^4$ configuration and Raman results under compressive strain along the [1 0 0] direction. \bf{}a\rm{}, $t_\mathrm{2g}$ orbital level splitting under tetragonal $\Delta$ and orthorhombic $\Delta_\mathrm{ort}$ crystal fields. $\Delta>0$ corresponds to flattened RuO$_6$ octahedra in Ca$_2$RuO$_4$, placing the $xz/yz$ orbitals above the $xy$ orbital. $\Delta_\mathrm{ort}>0$ reflects a shorter $a$-axis relative to the $b$-axis, resulting in a higher $az$ level than $bz$. \bf{}b\rm{}, Energy levels of $t_\mathrm{2g}^4$ electronic configuration including the crystal field and SOC effects. The higher energy $\tilde{J}$ = 2 states are not shown. The ground state singlet and the lowest doublet $T_{x/y}$ form an effective pseudospin $S$ = 1 basis. \bf{}c\rm{}, Measured magnon energies (circles) and theory (line) under compressive strain along the [1 0 0] direction. Circles with different colors denote different data sets. In our theory fits, $\Delta_{\rm ort}$ increases linearly with $\varepsilon$, while $\Delta = 250$ meV remains a constant.}
	\label{fig:3}
\end{figure}

\bigskip
\noindent{\large\textbf{Raman scattering}}\\
Fig.~\ref{fig:1}d presents Raman spectra of magnons and phonons under uniaxial strain along the [1 0 0] axis. The scattering geometry and polarization configurations are detailed in Methods. In the AFM state ($T$ = 25 K), unstrained Ca$_2$RuO$_4$ exhibits a sharp and intense magnon feature at 12.7 meV (3.07 THz) in the $B_\mathrm{1g}$ channel \cite{Sou17}, consistent with the magnon gap measured by inelastic neutron scattering \cite{Jai17}. Since the [1 0 0] axis corresponds to the shorter lattice constant, compression along this direction enhances the lattice orthorhombicity (Fig.~\ref{fig:1}b) and thus the magnetic anisotropy via pseudospin-lattice coupling \cite{Liu19}. Consequently, the magnon gap increases, as evidenced by the blue shift of the magnon peak. Both the magnon and phonon peaks exhibit nearly linear responses to strain, as summarized in Fig.~\ref{fig:3} and Supplementary Fig. 6, respectively.

A remarkable anomaly in the magnon spectrum is observed under compressive strain along the [1 1 0] direction (Fig. ~\ref{fig:1}e). At a critical strain $\varepsilon_{c} \sim$ 0.15\%, the magnon peak splits in the $B_\mathrm{1g}$ channel into two components: a dominant higher-energy peak and a weak lower-energy feature. Their combined spectral weight matches that of the original magnon (Supplementary Fig. 7). The dominant peak lies more than 10\% ($\sim$ 0.3 THz) above the pre-transition magnon energy, while the minor feature stays slightly below it. Both modes broaden and weaken rapidly upon warming (Supplementary Fig. 4), with a temperature dependence distinct from phonons yet analogous to the zero-strain magnon \cite{Sou17}. These characteristics confirm both sub-14 meV features above $\varepsilon_{c}$ as magnon excitations. The abrupt change in magnon spectrum signals a strain-driven magnetic phase transition near $\varepsilon_{c}$, as further examined by RXD (Fig.~\ref{fig:2}). The reversibility and reproducibility of this transition (Supplementary Fig. 7) confirms its intrinsic origin. 

\begin{figure}
	\centering{\includegraphics[width=0.48\textwidth]{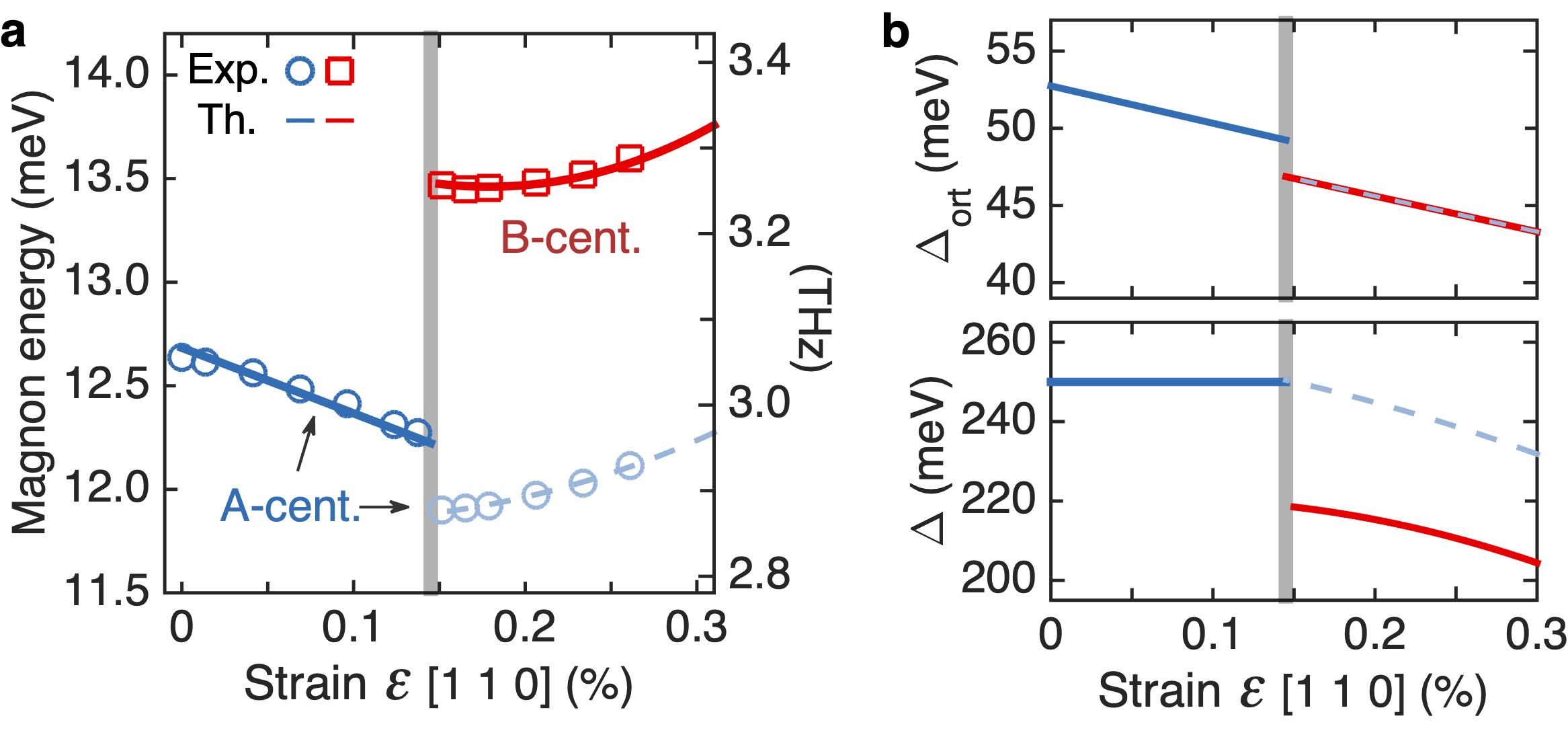}}
	\caption{\bf{} Raman results under the compressive strain along the [1 1 0] direction. a\rm{}, Measured magnon energies (blue circles: A-centered phase; red squares: B-centered phase) and theory (blue line: A-centered phase; red line: B-centered phase). Above the critical strain $\varepsilon_c \sim$ 0.15\% (vertical grey line), the light blue circles and dashed line represent the residual A-centered phase. \bf{}b\rm, Strain evolution of the crystal field parameters $\Delta_{\rm ort}$ (top) and $\Delta$ (bottom) obtained from the theory fits. Colors and line styles match those in (\bf{}a\rm{}). In the top panel, $\Delta_{\rm ort}$ decreases linearly with strain, with a slight drop at $\varepsilon_c$. The light blue dashed line (residual A-centered phase) overlaps the red solid line (B-centered phase) for $\varepsilon > \varepsilon_c$. In the bottom panel, $\Delta$ remains constant (250 meV) before the transition. At $\varepsilon_c$, an abrupt drop of $\Delta$ to $\sim 220$ meV occurs for the B-centered phase, while $\Delta$ remains continuous for the A-centered phase. After the transition, the strain dependence of $\Delta$ is described as $\Delta = \Delta_0 [ 1+a_1 (\varepsilon-\varepsilon_c)-a_2 (\varepsilon-\varepsilon_c)^2]$, with $a_1=0.22$, $a_2=1.4$ for the B-centered phase and $a_1=0.38$, $a_2=0.7$ for the residual A-centered phase, yielding the best fit to the data.}
	\label{fig:4}
\end{figure}

Fig.~\ref{fig:1}e also reveals a sharp contrast in the strain dependence of the magnon energy across $\varepsilon_{c}$. At low strain, the magnon peak exhibits a nearly linear red shift, whereas above $\varepsilon_{\mathrm{c}}$ the energies of both magnon features increase nonlinearly with strain. The reduction in lattice orthorhombicity under strain along the [1 1 0] direction (up to 0.22\%, Fig.~\ref{fig:1}c) explains the magnon gap decrease below $\varepsilon_{\mathrm{c}}$, but does not extend to the nonlinear increase above the transition. This distinct response suggests that mechanisms beyond orthorhombicity changes are required to describe the magnon behavior above the transition, as further elucidated by our theoretical analysis (Fig.~\ref{fig:4}).

We further observe that the opposite effects of compression along the [1 0 0] and [1 1 0] directions on lattice orthorhombicity (Fig.~\ref{fig:1}b-c) are reflected in the lattice dynamics: most phonon modes shift in opposite directions (Supplementary Fig. 3). An exception is the $A_\mathrm{g}$ phonon near 25.4 meV, which softens under both in-plane compression directions (Fig.~\ref{fig:1}d-e), suggesting that its eigenmode involves atomic vibrations along the expanding $c$-axis. Notably, the strain dependence of the mode frequency changes at $\varepsilon_{c}$ (Fig.~\ref{fig:1}e and Supplementary Fig. 6). Similar anomalies are also observed in other phonon modes, collectively suggesting concomitant structural modifications across the magnetic phase transition. 

No additional phonon modes appear across $\varepsilon_{c}$, indicating that the magnetic phase transition is not accompanied by a change in lattice symmetry.

\bigskip
\noindent{\large\textbf{Resonant x-ray diffraction}}\\
To investigate the magnetic phase transition under compressive strain along the [1 1 0] direction, we performed RXD measurements at the Ru-$L_3$ absorption edge on the same sample used for Raman spectroscopy. 

Fig.~\ref{fig:2}a-b shows the strain dependence of (1 0 2) and (1 0 3) Bragg reflections, which correspond to the ordering wave vectors for the A-centered and B-centered AFM structures, respectively. As illustrated in Fig.~\ref{fig:2}c, the two phases possess identical in-plane spin order but differ in their interlayer stacking at $z$ = 0.5. The pristine sample exhibits the A-centered phase, while the B-centered phase is typically stabilized by doping or pressure \cite{Bra98,Fri01,Ste05,Tan13,Pin18,Chi20,Por22}. With increasing strain along the [1 1 0] direction, the (1 0 2) reflection intensity drops as the (1 0 3) reflection rises from background, indicating a  A-to-B-centered phase transition near $\varepsilon \sim $ 0.15\% (Fig. \ref{fig:2}c). Moreover, this transition is reversible upon strain release (Supplementary Fig. 8). The strain-stabilized B-centered phase exhibits a higher N\'eel temperature ($T_\mathrm{N} \sim$ 135 K, Supplementary Fig. 9) than the A-centered phase ($T_\mathrm{N} \sim$ 113 K), consistent with previous reports \cite{Fri01,Por22,Ste05,Tan13}. 

The magnetic transition observed by RXD occurs at the same critical strain as identified by Raman spectroscopy, though the transition appears broader in the RXD data due to the large x-ray beam footprint on the sample (see Methods). Both techniques reveal a strain-driven magnetic phase transition accompanied by a pronounced modulation in the magnon gap. 

RXD measurements also reveal a residual (1 0 2) reflection that persists up to the maximal strain ($\varepsilon \sim $ 0.22\%), indicating a residual A-centered phase coexisting within the dominant B-centered phase. This is consistent with the post-transition Raman spectra showing both a weak and a strong magnon feature (Fig.~\ref{fig:1}e). We therefore assign the strong magnon peak to the majority B-centered phase and the weak one to the residual A-centered phase, supported by the close energy match between the weak peak and the pre-transition magnon. Under further strain, the dominant magnon peak grows more intense, in contrast to the weak one, which remains faint (Supplementary Fig. 7). 

\bigskip
\noindent{\textbf{\large{}Theoretical model}}\\
To describe the strain dependence of the magnon gap, we introduce an effective pseudospin model which takes into account spin-orbit and orbital-lattice coupling effects. 

In Ca$_2$RuO$_4$, the Ru$^{4+}$ ion contains four electrons in the $t_{2g}$ manifold, forming a low-spin state with total spin $s=1$ and effective orbital moment $l=1$. Spin-orbit coupling (SOC) $\lambda \vc s \cdot \vc l$ entangles these degrees of freedom into multiplets with total angular momentum $\tilde{J}$ = 0, 1 and 2, with $\tilde{J}$ = 0 being the ground state. Under the tetragonal crystal field $\Delta$, the excited $\tilde{J}$ = 1 triplet splits into a doublet $T_{x/y}$ at energy $E$ and a high-energy $T_z$ level, see Fig.~\ref{fig:3}a-b. Further, the orthorhombic crystal field $\tfrac{1}{2}\Delta_{\rm ort} (n_{az}- n _{bz})$, where $az = \frac{1}{\sqrt{2}} (yz+xz)$ and $bz= \frac{1}{\sqrt{2}} (yz-xz)$, splits the $T_{x/y}$ doublet into $T_{a/b}=\frac{1}{\sqrt{2}}(T_x\pm T_y)$ states by $\kappa$. 

The energies $E$ and $\kappa$, which quantify the level structure of the Ru$^{4+}$ ions, are determined by $\lambda$ and the crystal fields $\Delta$ and $\Delta_\mathrm{ort}$. 
We find that $E=\lambda \big( \tfrac{1}{2} +\sqrt{\tfrac{9}{4}-\delta+\delta^2} -\sqrt{1+\delta^2} \big)$, while the orthorhombic splitting $\kappa$, which is responsible for the in-plane magnetic anisotropy, is given by 
\begin{align}
\kappa= \frac{1}{2}  \bigg( 1-\frac{\delta}{\sqrt{1+\delta^2}}\bigg) \Delta_{\rm ort} \;,
\label{eq:1}
\end{align}
where $\delta=  \frac{\Delta}{2 \lambda}$. In the strong SOC limit ($\Delta\ll \lambda$), orbital moments remain unquenched and $\kappa \simeq \tfrac{1}{2}\Delta_{\rm ort}$; conversely, for large $\Delta$, orbital moments are quenched and $\kappa$ is suppressed as $(\lambda/\Delta)^2\Delta_{\rm ort}$.

The above level structure and magnetic excitations, within a Hilbert space comprising $\tilde{J}=0$ and $T_{a/b}$ states of the Ru$^{4+}$ ions, can be described in terms of the pseudospin $S=1$. Assigning $S_z=0$ state to the $\tilde{J}=0$ level, the minimal Hamiltonian for quasi-two-dimensional Ca$_2$RuO$_4$ can be cast in the following form \cite{Jai17}:
\begin{align}
\mathcal H = & J \sum_{\langle ij \rangle} (\vc S_i \cdot \vc S_j -\tau S_{zi} S_{zj}) + \sum_i \big[E S_{zi}^2 +\tfrac{1}{2} \kappa (S_{ai}^2 - S_{bi}^2)\big].
\label{eq:2}
\end{align}
Here, $J>0$ denotes the nearest-neighbour exchange coupling within the planes (the weak interlayer coupling $J_c \sim 10^{-2}$ meV \cite{Tre22,Kun15} has a negligible contribution to the magnon spectra \cite{Jai17} and thus not shown at this point). The exchange anisotropy with $\tau>0$ and single-ion anisotropy $E$ confine the magnetic moments to the $ab$-plane, resulting in $XY$ model type spin dynamics \cite{Tre22}. The orthorhombic anisotropy $\kappa$ term further aligns the moments along the $b$-axis, as observed in both the A-centered and B-centered phases \cite{Bra98,Fri01,Ste05,Pin18,Chi20}, and opens the magnon gap. 

The magnetic order in the model (equation~\eqref{eq:2}) arises from the condensation of the excited states, provided that the exchange interactions are strong enough to overcome the single-ion energy gap $E$~\cite{Kha13,Jai17}. The acoustic magnon gap, which is probed in the $B_{1g}$ channel, is given by the following equation~\cite{Jai17}:
\begin{align}
\omega = \sqrt{\kappa \big[ 4J (2-\tau) +\tfrac{1}{2} \tau E  + \kappa \big]} \;.
\label{eq:3}
\end{align}
This expression highlights the strong sensitivity of the magnon gap to the single-ion anisotropy $\kappa$ (equation~\eqref{eq:1}), which can be tuned by the crystal fields $\Delta$ and $\Delta_{\rm ort}$, and thus, by strain. In the following, we first analyze the magnon gap response to strain, and then discuss the magnetic phase transition induced by strain along the [1 1 0] direction.

\medskip
\noindent{\textbf{Strain along the [1 0 0] direction.}}
Compressive strain $\varepsilon$ along the [1 0 0] direction enhances the lattice orthorhombicity (Fig. \ref{fig:1}b). This should increase the orbital splitting $\Delta_\mathrm{ort}$ and hence the anisotropy parameter $\kappa$, see equation~\eqref{eq:1}. Neglecting possible small variations of $\delta$, we obtain a linear relation between $\kappa$ and strain $\varepsilon$:
\begin{align}
\kappa = \kappa_0 \bigg(1+\frac{\varepsilon}{\varepsilon_0}\bigg) \;,
\label{eq:kappa}
\end{align}
where $\varepsilon_0$ denotes the intrinsic orthorhombicity, and $\kappa_0=g\varepsilon_0$ is the corresponding in-plane anisotropy induced by the pseudospin-lattice coupling with strength $g$.

From equation~\eqref{eq:3} and equation~\eqref{eq:kappa}, the magnon gap is obtained as
\begin{align}
\omega \simeq \omega_0 \sqrt{1+\frac{\varepsilon}{\varepsilon_0}} \;,
\label{eq:4}
\end{align}
with $\omega_0 = \sqrt{\kappa_0 \big[4J(2-\tau)+\tfrac{1}{2}\tau E + \kappa_0\big]}$ being the intrinsic magnon gap.

Experimentally, $\omega_0 = 12.7$ meV and $\varepsilon_0=2.2\%$ are obtained for the strain-free crystal at 20 K. With only these measured parameters, equation~\eqref{eq:4} excellently reproduces the measured strain dependence of the magnon gap, see Fig.~\ref{fig:3}b. This consistency also supports the linear relation (equation~\eqref{eq:kappa}) between $\kappa$ and $\varepsilon$. 

Previous RIXS experiments have suggested that $\Delta= 250$ meV and $2\lambda=145$ meV ($\delta \simeq 1.7$) \cite{Gre19}, which give $E=27$ meV. Together with $J=5.8$ meV \cite{Jai17} and $\tau=0.5$, the intrinsic single-ion anisotropy $\kappa_0 = 3.6$ meV is evaluated from $\omega_0$. This gives the pseudospin-lattice coupling constant $g=\kappa_0/\varepsilon_0\sim 160$ meV. Based on equation~\eqref{eq:1} with $\delta \simeq 1.7$, one can deduce the intrinsic orthorhombic field $\Delta_{\rm ort}^0\simeq 53$ meV. 

\medskip
\noindent{\textbf{Strain along the [1 1 0] direction.}} Opposite to the previous case, compression along the [1 1 0] direction reduces the lattice orthorhombicity (Fig. \ref{fig:1}c). This should reduce the orbital splitting $\Delta_\mathrm{ort}$ and, according to equation~\eqref{eq:1}, decrease the in-plane anisotropy $\kappa$. Assuming again a linear relation between $\kappa$ and $\varepsilon$ at small strain values, we obtain the magnon gap as follows:    
\begin{align}
\omega \simeq \omega_0 \sqrt{1 - \frac{\varepsilon}{\varepsilon_0}} \;,\ \ \ {\rm when} \ \ \ \varepsilon  < \varepsilon_c \;.
\label{eq:omee2}
\end{align}
With the same values for $\omega_0$ and $\varepsilon_0$ quoted above, equation~\eqref{eq:omee2} agrees well with the experiment, see Fig.~\ref{fig:4}a. 

At the critical strain $\varepsilon_c \sim$ 0.15\%, the weak magnon peak associated with the residual A-centered phase softens by $\sim$ 0.3 meV; this can be attributed to a modest reduction of the orthorhombic field $\Delta_{\rm ort}$ by $\sim$ 2 meV at the phase transition [Fig.~\ref{fig:4}b (top panel)].  

The abrupt increase by $\sim 1.3$ meV of the magnon energy in the B-centered phase at $\varepsilon_c$ is unexpected, however, since strain along the [1 1 0] direction monotonically reduces the lattice orthorhombicity and thus decreases the $xz/yz$ orbital splitting $\Delta_{\rm ort}$. To understand this puzzling behavior, we recall that the in-plane magnetic anisotropy $\kappa$ (equation~\eqref{eq:1}) depends also on the tetragonal field $\Delta$. As discussed above (see also Supplementary Fig. 11), $\kappa$ strongly increases at smaller $\Delta$ values, reflecting the fact that SOC effects are enhanced in the cubic limit. We find that a drop of the tetragonal field $\Delta$ by $\sim30$ meV at $\varepsilon_c$ enhances $\kappa$ by $\sim 0.7$ meV, which translates into the magnon gap increase of $\sim 1.3$ meV as observed. 

We note that the abrupt change in crystal fields across $\varepsilon_c$ aligns with structural modifications that are evidenced by the concurrent change of the strain dependence of phonon frequency (Fig.~\ref{fig:1}e and Supplementary Fig. 6).

\begin{figure*}
	\centering{\includegraphics[width=1\textwidth]{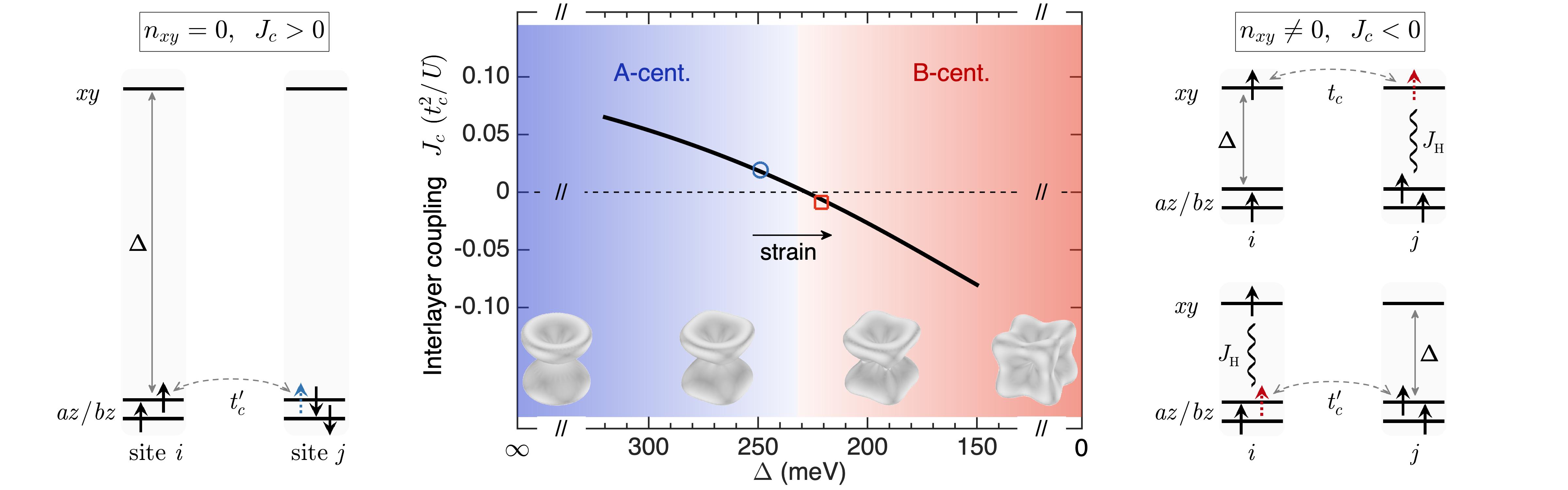}}
	\caption{\bf{}Calculated magnetic phase diagram and interlayer exchange processes.\rm{} \bf{}Middle\rm{}: Interlayer coupling $J_c$ (in units of $t_{c}^2/U$) as a function of tetragonal crystal field $\Delta$ (solid black curve), where $J_c >$ 0 corresponds to AFM interlayer coupling. Under compressive strain along the [1 1 0] direction, $J_c$ changes sign as $\Delta$ decreases abruptly from 250 (blue circle) to 220 (red square) meV across the critical strain $\varepsilon_c$, driving the transition from A-centered to B-centered phase. At the bottom, spin-orbit wave functions for the two-hole $t_\mathrm{2g}$ configuration are visualized at different values of $\Delta$ with $2\lambda=145$ meV. \bf{}Left\rm: In the hole language where each site hosts two spins (black arrows), the schematic visualization of AFM exchange between interlayer nearest-neighbor sites $i$ and $j$ at large $\Delta$ is shown. The blue arrow indicates a spin hopping between sites, aligned antiparallel to the existing spins. \bf{}Right\rm: At small $\Delta$ with the finite $xy$ hole density, two distinct hopping processes (dashed double arrows) contribute to ferromagnetic coupling via Hund's coupling $J_\mathrm{H}$. The red arrow denotes the spin hopping parallel to the existed spins. Here, $t_{c}$ ($t_{c}^\prime$) denotes interlayer hopping between $xy$ ($az/bz$) levels. In our model, $t_{c} =2.4 t_{c}^\prime$.}
	\label{fig:5}
\end{figure*}

\medskip
\noindent{\textbf{Magnetic phase transition.}} Interestingly, the strain-induced reduction of the tetragonal crystal field $\Delta$ has a decisive impact on the sign of interlayer coupling $J_c$, offering thereby a natural explanation to the transition from the A-centered to the B-centered magnetic structure. Microscopically, the exchange processes that determine the $J_c$ values depend on the spin-orbital composition of the Ru electrons, which in turn is decided by competition between the crystal fields and SOC. 

We have considered the interlayer exchange processes involving $t_{2g}$ orbitals and Hund's coupling effects (see Supplementary Note 3B for details). The calculated dependence of the interlayer coupling $J_c$ on a tetragonal field $\Delta$ is shown in the middle panel of Fig.~\ref{fig:5}. At large $\Delta$, the ground-state wave function carries negligible $xy$-orbital weight ($n_{xy}\sim 0$), and interlayer exchange is dominated by $az/bz$ orbital hopping processes resulting in AFM coupling $J_c>0$ (left panel of Fig.~\ref{fig:5}). By contrast, when $\Delta$ is reduced under strain, the spin-orbit entangled wave function acquires substantial $xy$-orbital character. This change activates the $xy$-orbital hopping channels, which favor ferromagnetic interactions $J_c<0$ through Hund's coupling (right panel of Fig.~\ref{fig:5}). Thus, the strain-driven modification of the wave functions increases the $xy$-hole density (Supplementary Fig. 13 and Note 3B), which reverses the sign of the interlayer coupling $J_c$ and drives the magnetic phase transition from A-centered to B-centered structure. 

\bigskip
\noindent{\textbf{\large{}Discussion}}\\
The sign reversal of the interlayer coupling $J_c$ is accompanied by a pronounced change in the magnon gap of $\sim$ 0.3 THz---two orders of magnitude larger than the $J_c$ value ($\sim 4\times10^{-3}$ THz \cite{Tre22,Kun15}) itself. This dramatic effect originates from the modulation of the spin-orbit entangled wave functions by strain.

In our experiment, the strain-stabilized B-centered phase emerges with reduced lattice orthorhombicity and an expanded $c$-axis, consistent with earlier reports on doped and pressurized Ca$_2$RuO$_4$ \cite{Bra98,Chi20,Pin18,Fri01,Por22,Ste05}. Our approach achieves this magnetic transition with significantly smaller structural modifications, as summarized in Supplementary Table 2. The low critical strain highlights the potential of strain engineering for controlling magnetic states in Ca$_2$RuO$_4$. 

With slight modification of the Ca$_2$RuO$_4$ lattice structure, either by dilute chemical substitution or by depositing thin films on suitable substrates, it should be possible to tune the critical strain close to zero, thus maximizing the strain sensitivity of the magnon energy in analogy to superconducting transition-edge sensors for ultra-sensitive cryogenic photon or particle detection \cite{Luc24}. When the transition is triggered and the magnon energy is thus enhanced in a THz magnon conduit, it becomes impassable for magnons injected from a strain-free region. Such an arrangement will thus function as an ultrasensitive magnetic switch. Due to the extreme strain sensitivity near the magnetic phase transition, one also expects large nonlinearities, as strain fields associated with a given magnon can affect the propagation of other magnons. Such nonlinearities may enable further transition-edge device functionalities. 

In summary, our study reveals a pronounced modulation of the magnon gap at a strain-controlled magnetic phase transition in Ca$_2$RuO$_4$ and provides a quantitative explanation of this effect in terms of in-situ control of the effective spin-orbit coupling on the Ru ions. The extreme sensitivity of both the magnetic ground state and the elementary excitations to external strain offers tantalizing prospects for ultrasensitive THz magnonic devices. 


\clearpage
\newpage

\noindent{\large\textbf{Online Methods}}\\

\noindent\textbf{Sample and uniaxial strain}

\noindent High-quality single crystals of Ca$_2$RuO$_4$ with $T_\mathrm{N}$ = 113 K were grown via the optical floating-zone technique. Twin-free crystals were identified and selected through Laue diffraction and magnetization measurements. To apply uniaxial strain, the selected crystals were cut into needle-like specimens with dimensions listed in Supplementary Table S1. Then, each specimen was mounted onto a custom-designed sample carrier using epoxy Stycast 2850FT, and fixed onto the strain device with screws, as shown in Fig.~\ref{fig:1}a. Uniaxial strain was applied by controlling the voltage of the piezoelectric device, while monitoring the displacement in situ with an internal capacitor gauge and an ultra-high-sensitivity capacitance bridge.

Prior to Raman measurements, the strain $\varepsilon$ was calibrated by determining the lattice constants via X-ray diffraction on the same sample at matching temperatures (see Supplementary Note 1 for more details). The pristine lattice constants (space group $Pbca$), measured at 20 K, are $\it{a}\rm{}_0$ = 5.382 $\rm{}\AA$, $\it{b}\rm{}_0$ = 5.622 $\rm{}\AA$ and $\it{c}\rm{}_0$ = 11.728 $\rm{}\AA$, yielding an intrinsic lattice orthorhombicity of $\varepsilon_0=(b_0-a_0)/(b_0+a_0)=2.2\%$. To model the strain dependence of the zone-center magnon energy which is closely linked to the change in lattice orthorhombicity, we introduce an effective strain $\varepsilon$ defined as $|(b-a)/(b+a)-\varepsilon_0|$, where $a$ and $b$ are lattice constants under strain.\\ 
\noindent\textbf{Raman scattering}

\noindent Raman scattering measurements were performed using a Jobin Yvon LabRAM HR800 spectrometer. The sample with strain device was mounted in a closed cycle cryostat under ultra-high vacuum, and measured in a confocal backscattering geometry at around 25 K. Using the 632.8 nm excitation line with a power of $\sim$ 1 mW, the laser beam was focused to a 5-$\mu m$-diameter spot on the sample surface. The strain-dependent spectra were taken on the same spot within a strain cycle. To clearly resolve the shifts of Raman features under strain, the scattered photons were resolved by a 1800 grooves/mm grating with an energy resolution of $\sim$ 0.04 meV. 

The Raman spectra presented in the maintext are taken with the linear polarization configurations for incident and scattered photons. Since the sample orientation in our strain setup is fixed with respect to the Raman spectrometer, the linear polarization of incident photons has to be either parallel or perpendicular to the strain direction. For strain along the [1 0 0] direction, linear polarization configurations $aa$ (parallel polarization) and $ab$ (crossed polarization) detected the $A_\mathrm{g}$ and $B_\mathrm{1g}$ symmetry channels separately; whereas for strain along the [1 1 0] direction, both symmetries were mixed in the parallel polarization $xx$ ($x$ is along the [1 1 0] direction), while the crossed linear polarization configuration $xy$ ($y$ is perpendicular to $x$) detected no Raman features. See Supplementary Fig. 3 for the sketch of polarizations and more detailed data. Despite this mixing for the measurements using linear polarizations, magnon features below 14 meV remain clearly distinguishable from phonons by their distinct energy scales. Additional measurements using circular polarizations $LL$ and $RL$ were performed to disentangle the $A_\mathrm{g}$ and $B_\mathrm{1g}$ symmetry channels, confirming that the split magnon features below 14 meV after the phase transition belong to the $B_\mathrm{1g}$ symmetry (Supplementary Fig. 5).\\

\noindent\textbf{Resonant x-ray diffraction}

\noindent Resonant x-ray diffraction (RXD) measurements were performed at the P09 beamline of the PETRA-III synchrotron at the Ru $L_3$ edge (2839.6 eV) in a reflection scattering geometry. The same sample used for Raman measurements were examined under compressive strain along the [1 1 0] direction. The strain device was mounted in a closed-cycle cryostat on a four-circle diffractometer. The incident x-rays were linearly polarized ($\sigma$-incident), but no polarization analysis was performed on the scattered beam. The beam profile was around 200 $\times$ 80 $\mu m^2$, and the vertical footprint on the sample was elongated at non-normal incidence during measurements. This large beam spot averaged over spatial strain inhomogeneities and resulted in the broadening of magnetic transition compared to that revealed by Raman measurements. 

\bibliographystyle{naturemag}
\bibliography{Reference}

\bigskip
\bigskip
\noindent{\textbf{Acknowledgements}}\ We acknowledge support from the European Research Council under Advanced Grant No. 101141844 (SpecTera). L.W. and S.H. are partially supported by the Alexander von Humboldt foundation. H. L. acknowledges the support by the National Natural Science Foundation of China (Grant No. 12574066) and the Fundamental Research Funds for the Central Universities (Grant No. KG202501). For RXD measurements, we acknowledge DESY (Hamburg, Germany) for providing experimental facilities and allocating beamtime (the Long Term Proposal II-20200008) at the P09 beamline of PETRA-III. We would like to further thank the technical assistance from J. Burkhardt, S. Mayer, H. Gretarsson, P. Wochner and B. Bruha, and the assistance in initializing the project from J. Bertinshaw. We also thank C. Hicks and A. Mackenzie for designing and providing the uniaxial strain device and for fruitful discussions. 

\bigskip
\noindent{\textbf{Author contributions}}\ L.W., M.M., and B.K. conceived the research project and designed the experiments. M.K. and L.W. grew and characterized the single crystals. L.W. prepared the sample for uniaxial-strain application and performed Raman measurement. L.W., S.H., Y.L., M.K., K.H., and S.F. preformed RXD measurement. L.W. analyzed the Raman and RXD data, with the input from G.K., M.M., and B.K.. H.L. and G.K. performed the theoretical analysis. L.W., H.L., G.K., and B.K. wrote the paper with contributions from all authors.

\clearpage                
\onecolumngrid            
\noindent {\it Supplementary Information} accompanies this paper at the ancillary files.
\twocolumngrid  

\end{document}